\documentclass[a4paper]{jpconf}
\usepackage{graphicx}
\usepackage{amsmath}
\usepackage{bbold}
\begin{document}
\title{Finite-range separable pairing interaction in Cartesian coordinates }

\author{A.M. Romero$^{1}$, J. Dobaczewski$^{1,2,3}$, A. Pastore$^{1}$}

\address{$^1$Department of Physics, University of York, Heslington, York YO10 5DD, United Kingdom\\
         $^2$Institute of Theoretical Physics, Faculty of Physics, University of Warsaw, ul. Pasteura 5, PL-02-093 Warsaw, Poland\\
$^3$Helsinki Institute of Physics, P.O. Box 64, FI-00014 University of Helsinki, Finland}

\ead{amr560@york.ac.uk}

\begin{abstract}
Within a simple SO(8) algebraic model, the coexistence between
isoscalar and isovector pairing modes can be successfully described
using a mean-field method plus restoration of broken symmetries. In
order to port this methodology to real nuclei, we need to employ
realistic density functionals in the pairing channel. In this
article, we present an analytical derivation of matrix elements of a
separable pairing interaction in Cartesian coordinates and we correct
errors of derivations available in the literature. After
implementing this interaction in the code {\sc hfodd}, we study evolution
of pairing gaps in the chain of deformed Erbium isotopes, and we
compare the results with a standard density-dependent contact pairing
interaction.
\end{abstract}

\section{Introduction}

Pairing correlations play a crucial role in understanding nuclear
phenomena, such as, for example, the odd-even mass
staggering~\cite{(Bro13)}. Due to residual pairing interactions,
fermions close to the Fermi energy tend to form a condensate of
Cooper pairs that are the source of superfluidity within the BCS
model~\cite{(Bar57)}. Since in a nucleus we have two types of
particles, there can appear, in principle, both isoscalar and isovector Cooper pairs.
There is substantial experimental evidence concerning the nuclear
superfluidity arising from the isovector pairing, however, a direct
observation of the isoscalar proton-neutron condensate remains
elusive~\cite{(Fra14)}.

In a recent article~\cite{(Rom19a)}, using the mean-field
approximation~\cite{ben03} applied to a simple SO(8)
model~\cite{(Pan69),(Kot06)}, we showed that it is possible to obtain
the coexistence of the two types of condensate. The main conclusion
of our work is that the crucial ingredient to observe such a
coexistence is to apply the variational principle for a projected
(symmetry-restored) mean-field states. This encouraging result
motivated us to implement the same technique within a realistic
nuclear density-functional theory (DFT). The first step in this
direction is to choose a realistic pairing functional that is capable
of reproducing basic properties of the isovector pairing and at the
same time allows for opening the isoscalar pairing channel. To this
end, in this article we discuss derivations and implementations
related to the finite-range separable pairing
interaction~\cite{dug04,tia09}. The advantage of using such a pairing
force is that it is free from ultraviolet divergences~\cite{bul02}
and at the same time it requires computational effort that is
comparable with a simple density dependent delta
interaction~\cite{ber91}.

The article is organised as follows. In Section~\ref{sec:SO8}, we
briefly summarise the main findings of Ref.~\cite{(Rom19a)}.
In Section~\ref{sec:separab}, we present derivation of matrix
elements of the separable interaction in Cartesian coordinates and
in Section~\ref{sec:results} we show sample results obtained for pairing gaps.
Finally, we present our conclusions in Section~\ref{sec:concl}.

\section{The SO(8) model}\label{sec:SO8}

\begin{figure}[h]
\begin{center}
\begin{minipage}{16pc}
\includegraphics[width=16pc]{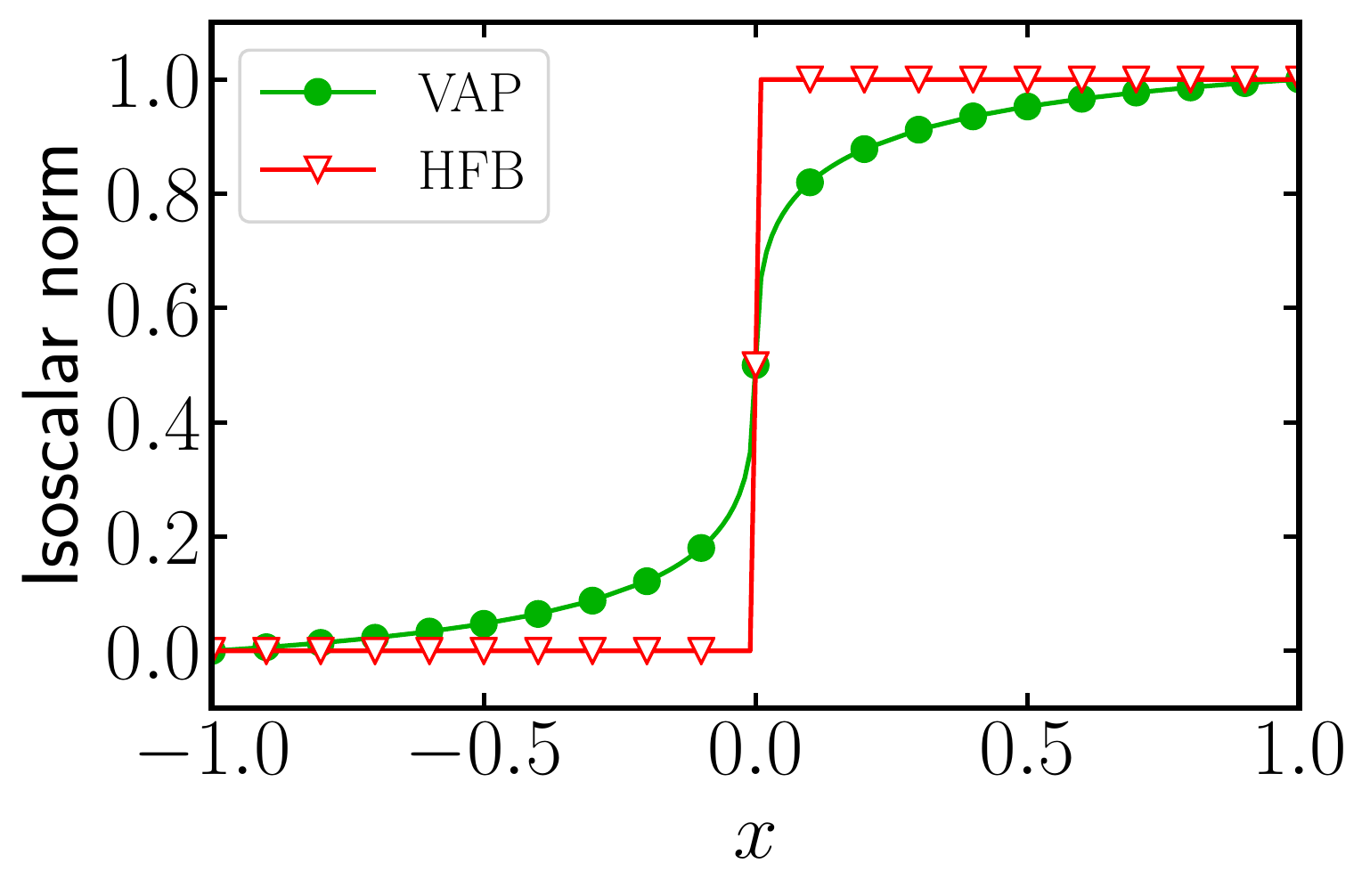}
\caption{\label{figSO8}Evolution of the isoscalar norm as a function
of the mixing parameter $x$ in Eq.~(\ref{eq:ham}), obtained in Ref.~\cite{(Rom19a)} using the HFB (open
symbols) and VAP methods (full symbols).}
\end{minipage}\hspace{2pc}%
\begin{minipage}{16pc}
\includegraphics[width=14pc]{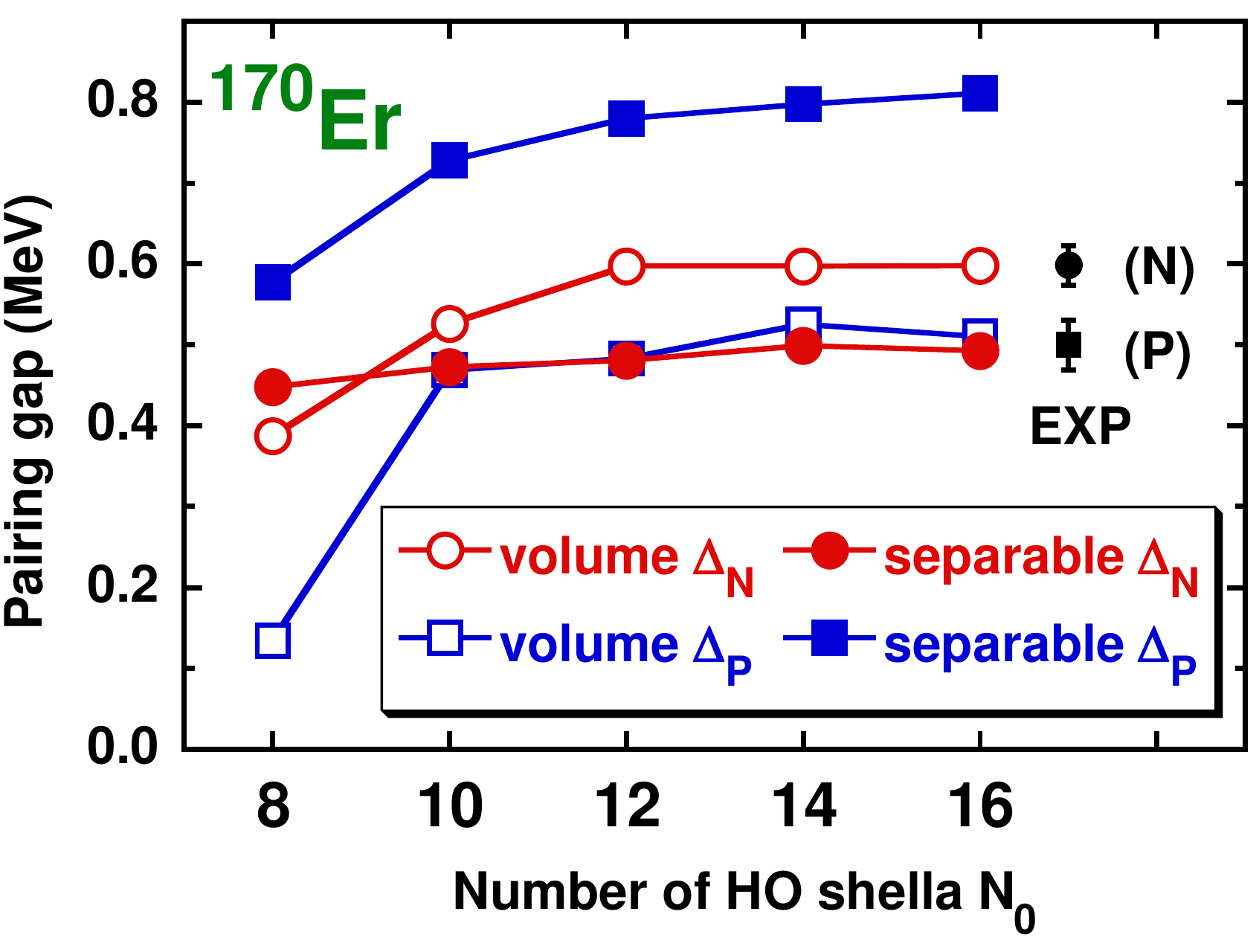}
\caption{\label{Convergence} Convergence of the neutron and proton
pairing gaps in $^{170}$Er with increasing number of the HO shells
$N_0$, determined for the zero-range volume and separable pairing
interaction.}

\end{minipage}
\end{center}
\end{figure}

The SO(8) model is based on a simple Hamiltonian written as a
linear combination of terms that depend on the isovector (isoscalar) pair creation
$\hat{P}^+(\hat{D}^+)$ and annihilation operators $\hat{P} (\hat{D})$,
\begin{equation}\label{eq:ham}
\hat{H} = - g(1-x) \sum_{\nu=0,\pm1} \hat{P}^+_{\nu} \hat{P}_{\nu}
          - g(1+x) \sum_{\mu=0,\pm1} \hat{D}^+_{\mu} \hat{D}_{\mu},
\end{equation}
where the sums run over all possible projections of the isospin
(spin) $\nu$ ($\mu$), $g$ is the pairing strength and $x$ is a mixing
parameter controlling the relative competition between isoscalar and
isovector contributions. By inserting Eq.~(\ref{eq:ham}) into the
Hartree-Fock-Bogoliubov (HFB) equations, we can observe the evolution
of the pairs as a function of the mixing parameter $x$, as shown in
Fig.~\ref{figSO8}. It is seen that, within the HFB
approximation, only full isovector or full isoscalar pair condensates can
exist, with a sharp transition between these two regimes that occurs at $x=0$.

The mean-field HFB state that minimizes the average value of
Hamiltonian (\ref{eq:ham}) breaks symmetries of the Hamiltonian such
as spin, isospin, and particle-number. The main thrust of
Ref.~\cite{(Rom19a)} was to apply projection techniques to select a
component of the HFB state with good quantum numbers and then apply
the variational principle to find the minimum of the corresponding
average energy. Such a procedure is called \emph{variation after
projection} (VAP).

For the VAP states, the coexistence of both types of pairs is
obtained for all values of $x$, Fig.~\ref{figSO8}, apart from the
two limiting cases of purely isovector ($x=-1$) or purely isoscalar
($x=1$) Hamiltonian. We tested the validity of the VAP approach by
comparing our results with the exact ones. For various values of
particle-number, spin, and isospin, we obtained a remarkable
agreement of the corresponding ground state energies. The relative
deviations turned out to be always below $1.5\%$. We also obtained a
perfect agreement between the VAP and exact deuteron-transfer matrix
elements. We refer to our article \cite{(Rom19a)} for details.

\section{Separable pairing interaction}\label{sec:separab}

In this Section, we present a detailed derivation of the matrix
elements of the separable interaction in a Cartesian basis. The
separable interaction has the general form
\begin{equation}\label{sepint}
\begin{split}
\hat{V}(\mathbf{r_1},\mathbf{r_2};\mathbf{r_1}',\mathbf{r_2}')
&= -\delta(X-X')\delta(Y-Y')\delta(Z-Z') P(x)P(y)P(z)P(x')P(y')P(z')\\
&\times[\tilde{W} \hat{\mathbb{1}} + \tilde{B} \hat{P}^{\sigma} - \tilde{H} \hat{P}^{\tau} - \tilde{M} \hat{P}^{\sigma} \hat{P}^{\tau} ]\;,
\end{split}
\end{equation}
where $\mathbf{r_i}=(x_i,y_i,z_i)$, $x=x_1-x_2$ and
$X=\frac{1}{2}(x_1+x_2)$ are the relative and center-of-mass
coordinates (equivalent symbols hold for the other Cartesian directions). For
the form factor, we chose a simple Gaussian as
\begin{equation}\label{formfactor}
P(x) = \frac{1}{\sqrt{\pi}a} e^{-x^2/a^2}\;,
\end{equation}
where $a= 1.232$\,fm is the range~\cite{ves11}. Symbols $\tilde{W}$, $\tilde{B}$, $\tilde{H}$, and $\tilde{M}$
denote adjustable parameters (Wigner, Bartlett, Heisenberg, and Majorana
coupling constants), $\hat{\mathbb{1}}$ is the identity operator, and
$\hat{P}^{\sigma}$ and $\hat{P}^\tau$ are spin and isospin exchange
operators, respectively. Because form factor (\ref{formfactor})
is symmetric in space coordinates, only parameters $\tilde{W}$ and $\tilde{B}$ are independent.

To calculate the matrix elements of interaction (\ref{sepint}),
we define the basis of two particles coupled to total spin $S=0$ as
\begin{equation}
|n_1n_2,S=0\rangle = \phi_{n_1}(x_1,b)\phi_{n_2}(x_2,b) |S=0\rangle\;,
\end{equation}
where $\phi_{n}(x,b)$ is the harmonic oscillator (HO) wavefunction with quantum number $n$ and
oscillator constant $b$,
\begin{equation}\label{eq:HO}
\phi_n (x,b) = b^{1/2} H_n^{(0)}(bx) e^{-b^2x^2/2}\;.
\end{equation}
In Eq. (\ref{eq:HO}), we used a normalized version $H_n^{(0)}(x)$ of the Hermite polynomials $H_n(x)$, defined as
\begin{equation}
H_n^{(0)}(x) = (\sqrt{\pi}2^n n!)^{-1/2} H_n(x)\;.
\end{equation}
We now evaluate the matrix elements in the new basis as
\begin{equation}\label{eq:mel}
\langle n_1'n_2',S=0|\hat{V}|n_1n_2,S=0\rangle = -I_x I_y I_z (\tilde{W}-\tilde{B})\;,
\end{equation}
where  $I_x$ is defined as
\begin{equation}
I_x(n_1n_2;n'_1n'_2) = \int dx_1 dx_2 dx'_1 dx'_2 \delta(X-X') P(x) P(x') \phi_{n_1}(x_1,b)\phi_{n_2}(x_2,b) \phi_{n'_1}(x'_1,b)\phi_{n'_2}(x'_2,b)\;,
\end{equation}
and similarly for $I_y$ and $I_z$. To calculate $I_x$, it is
convenient to express the two-particle basis in the center of mass
coordinates,
\begin{equation}\label{eq:moshtrans}
\phi_{n_1}(x_1,b)\phi_{n_2}(x_2,b) = \sum_{nN} M_{n_1n_2}^{nN} \phi_{n}(\tilde{x},b)\phi_{N}(\tilde{X},b)\;,
\end{equation}
where $\tilde{x} =\frac{x}{\sqrt{2}}$, $\tilde{X} =\sqrt{2}X$, and
symbol $M_{n_1n_2}^{nN} = \langle n_1 n_2 | nN\rangle$ denotes the
Moshinsky coefficients, which are given by the
expression~\cite{dobaczewski2009solution}
\begin{equation}
M_{n_1n_2}^{nN} = \frac{\sqrt{n_1!n_2!n!N!}}{2^{\frac{n+N}{2}}}  \delta_{n_1+n_2,n+N} \sum_{k=\max(0,n_1-n)}^{\min(N,n_1)}  \frac{(-1)^{n-n_1+k}}{k!(N-k)!(n_1-k)!(n-n_1+k)!}\;.
\end{equation}
We now obtain a new expression for $I_x$,
\begin{equation}\label{eq:integrale}
\begin{split}
I_x(n_1n_2;n'_1n'_2) = &\sum_{n,n',N,N'} M_{n_1n_2}^{nN} M_{n'_1n'_2}^{n'N'} \\ & \times \int dx dx' dX dX' \delta(X-X') P(x) P(x') \phi_{n}(\tilde{x},b)\phi_{n'}(\tilde{x}',b) \phi_{N}(\tilde{X},b)\phi_{N'}(\tilde{X'},b)\;,
\end{split}
\end{equation}
The integral over the coordinates $X,X'$ can be performed analytically, and gives
\begin{equation}
\begin{split}
& \int dX dX' \delta(X-X') \phi_{N}(\tilde{X},b)\phi_{N'}(\tilde{X'},b)  = \frac{1}{\sqrt{2}}\delta_{NN'}\;,
\end{split}
\end{equation}
where we used the normalization condition of the Hermite
polynomials $H_n^{(0)}(x)$~\cite{abramowitz1965handbook}.
Inserting this result into Eq.~(\ref{eq:integrale}), we get
\begin{equation}\label{eq:melnocom}
I_x(n_1n_2;n'_1n'_2) = \frac{1}{\sqrt{2}}\sum_{n,n',N} M_{n_1n_2}^{nN} M_{n'_1n'_2}^{n'N} \int dx dx' P(x) P(x') \phi_{n}(\tilde{x},b)\phi_{n'}(\tilde{x}',b)\;,
\end{equation}
which can be written as
\begin{equation}
I_x(n_1n_2;n'_1n'_2) = \frac{1}{\sqrt{2}}\sum_{n,n',N} M_{n_1n_2}^{nN} M_{n'_1n'_2}^{n'N} W(n)W(n')\;,
\end{equation}
where $W(n)=\int dx P(x) \phi_{n}(\tilde{x},b)$. Using the properties
of the HO wavefunction, this integral can be expressed as
\begin{equation}
W(n)  = \frac{\sqrt{2}}{\sqrt{\pi ba^2}}\int dt e^{-2t^2/(ba)^2} H_{n}^{(0)}(t) e^{-\frac{t^2}{2}}\;,
\end{equation}
where we changed the integration variable to $t=bx/\sqrt{2}$. Then using the following identity
\begin{equation}\label{eq:hermid}
\int_{-\infty}^{+\infty} du H_{2m}^{(0)}(\alpha u) e^{-u^2} = \pi^{1/4} \frac{\sqrt{(2m)!}}{m!} \left(\frac{\alpha^2-1}{2}\right)^m\;,
\end{equation}
we obtain the final result
\begin{equation}
W(n)  =  \frac{2\pi^{-1/4}\sqrt{b}}{\sqrt{a^2b^2+4}}  \frac{\sqrt{n!}}{(n/2)!} \bigg(\frac{a^2b^2-4}{2a^2b^2+8}\bigg)^{n/2}\;.
\end{equation}
We observe that the sum over $n,n'$ in Eq.~(\ref{eq:melnocom}) can be
further reduced by using the fact that the Moshinsky coefficients
will be zero unless $n=n_1+n_2-N$ and $n'=n'_1+n'_2-N$. It is
convenient to rewrite the final expression of the matrix elements as
\begin{equation}\label{eq:totmatsep}
I_x(n_1n_2;n'_1n'_2) = \frac{1}{\sqrt{2}} \sum_{N=0}^{n_1+n_2} G(N,n_1,n_2) G(N,n'_1,n'_2) \;,
\end{equation}
where
\begin{equation}
G(N,n_1,n_2) = M_{n_1n_2}^{n_1+n_2-N,N} W(n_1+n_2-N)\;.
\end{equation}

It is worth noting that a similar derivation of matrix elements in
Cartesian space has already been given in Ref.~\cite{nik10}, however,
in their derivation we have found several errors and missing factors
in the final expressions. In addition, an alternative derivation of
the analogous matrix elements of more general separable interactions
was given in Ref.~\cite{rob10}.

\section{Pairing gaps}\label{sec:results}

We tested our derivations and implementation in the code {\sc hfodd}
(v2.91a)~\cite{(Sch17b),(Dob19)} by comparing our results with the
ones obtained by the code {\sc hosphe}~\cite{car10} -- an HFB solver in
spherical symmetry, which also contains an implementation of the same
separable interaction in the pairing channel. For a fixed number of
shells, we reproduced the results of Fig.~1 of Ref.~\cite{ves11} up
to an eV accuracy, thus obtaining a very strong test on the
correctness of our results.

In this work, we present an example of calculations performed for the
chain of deformed Erbium isotopes. First, using the SLy4
functional~\cite{cha97} in the particle-hole channel, we roughly
adjusted parameter $\tilde{W}=-\tilde{B}=-300$\,MeV\,fm$^3$, Eq.~(\ref{eq:mel}), to reproduce
the values of experimental neutron, $\Delta_N$, and proton,
$\Delta_P$, pairing gaps in $^{170}$Er. In Fig.~\ref{Convergence},
we plotted the calculated $^{170}$Er neutron and proton pairing gaps as
functions of the number of the HO shells $N_0$. We observe that the
results converge nicely as a function of $N_0$, and we can consider
that at $N_0=14$ the pairing gaps are sufficiently
converged~\cite{ben00}. We note that the charge-symmetric separable
pairing interaction used here is not capable of reproducing the
experimental values of the $^{170}$Er neutron and proton pairing gaps
simultaneously.

In the same figure, we also report the analogous
convergence obtained for a simple charge-symmetric volume contact
pairing interaction adjusted to the same experimental data, which for
the equivalent-spectrum cut-off of 60\,MeV gave the strength of
$V_0=-195$\,MeV\,fm$^3$. Contrary to typical applications of the
volume pairing, where different strengths are used for neutrons and
protons, cf.~Ref.~\cite{(Kor10)}, here the experimental values of the $^{170}$Er
neutron and proton pairing gaps are perfectly reproduced by the
charge-symmetric parametrization.

\begin{figure}[h]
\begin{center}
\includegraphics[width=30pc]{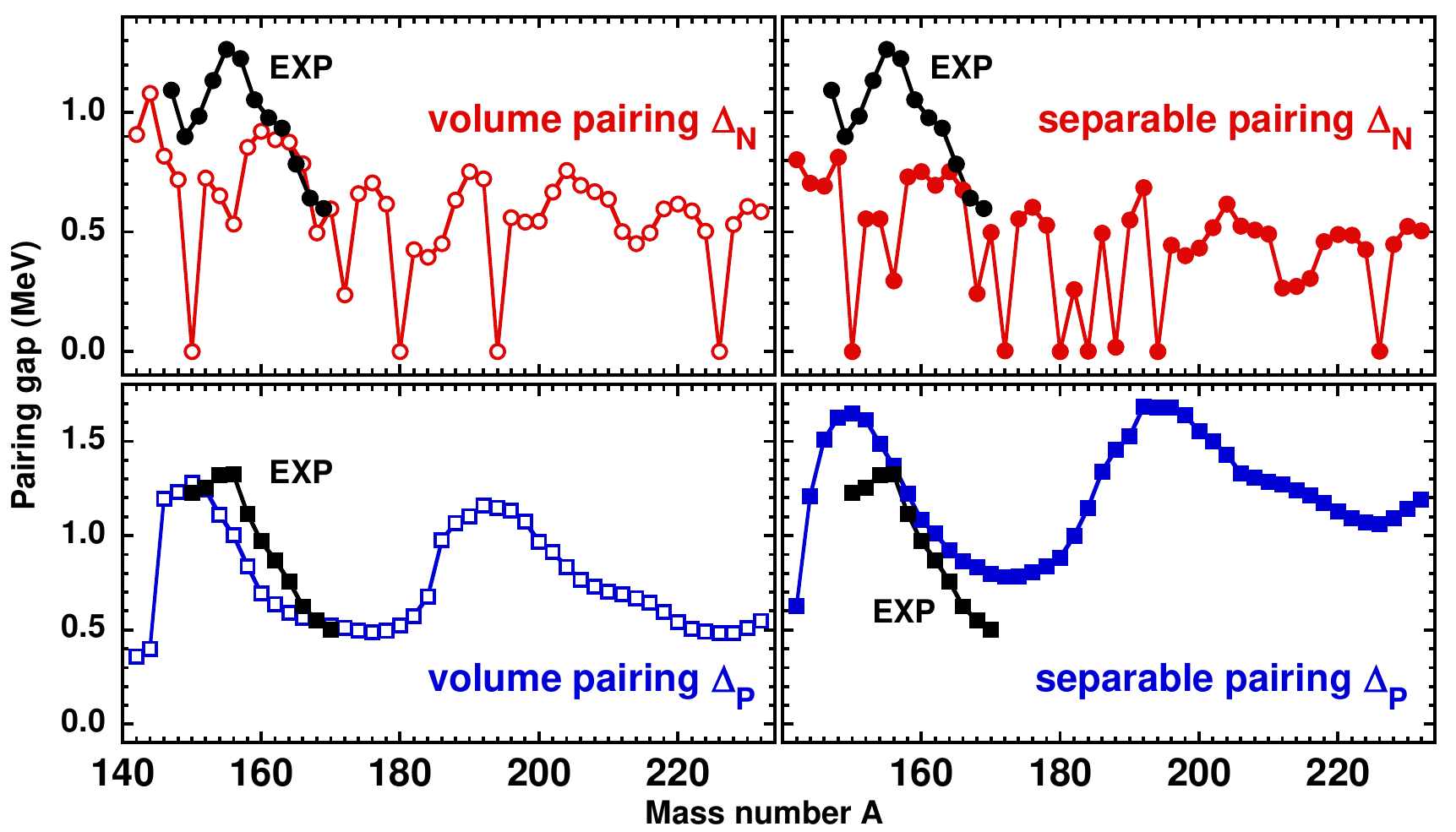}
\caption{\label{Pairing} Calculated neutron (upper panels) and proton
(lower panels) pairing gaps in Erbium isotopes determined for the
zero-range volume (left panels) and separable (right panels) pairing
interaction for $N_0=14$ HO shells. Neutron and proton experimental
values are calculated as averages of the three-point mass staggering
centered on two adjacent odd isotopes and isotones, respectively.}
\end{center}
\end{figure}

Having adjusted the strengths of the two pairing interactions, in
Fig.~\ref{Pairing} we compare the isotopic behaviour of the
calculated pairing gaps in Erbium isotopes. We observe that the
separable pairing tends to give a weaker pairing for neutrons and a
stronger for protons than the volume pairing. It is interesting to
note that in some nuclei the separable interaction leads to a
vanishing neutron gap, whereas the volume interaction may still give
there a non-zero value of $\Delta_N\approx 0.5$ MeV. Since Erbium nuclei
are deformed (apart from the semi-magic isotopes), we checked that
the values of the quadrupole moment, obtained for both interactions,
are practically the same.

\section{Conclusions}\label{sec:concl}

In this contribution, we briefly discussed the question of how to
observe and describe, within a simple SO(8) model, the coexistence of
isoscalar and isovector pairs by employing mean-field wave functions
with all relevant broken symmetries restored. In the perspective of
applying such a procedure to the realistic cases of finite nuclei,
we presented a detailed derivation of matrix elements of a separable
finite-range interaction in Cartesian basis, where we corrected
errors found in the literature derivations. To illustrate a practical
implementation in the 3D code {\sc hfodd}, we also presented a series of
systematic HFB calculations in Erbium isotopes, and we compared
the results with those obtained using a simpler zero-range volume
interaction.

\ack

This work was partially supported by the STFC Grants No.~ST/M006433/1
and No.~ST/P003885/1, and by the Polish National Science Centre under
Contract No.~2018/31/B/ST2/02220.
We acknowledge the CSC-IT Center for Science Ltd., Finland, for the allocation of
computational resources.

\section*{References}
\bibliographystyle{iopart-num}

\bibliography{Antonio.INPC2019-15}

\end{document}